\documentclass[global,referee]{svjour}
\usepackage{graphics}
\usepackage{epsfig}
\journalname{myjournal}
\begin{document}
\title{Full Counting Statistics of Multiple Andreev Reflections in
incoherent diffusive superconducting junctions}
\author{P. Samuelsson 
}                     
\offprints{}          
\institute{Division of Mathematical Physics, Lund University, S\"olvegatan 14 A, S-223 62 Lund, Sweden}
\date{Received: date / Revised version: date}
%
\maketitle
\begin{abstract}
We present a theory for the full distribution of current fluctuations
in incoherent diffusive superconducting junctions, subjected to a
voltage bias. This theory of full counting statistics of incoherent
multiple Andreev reflections is valid for arbitrary applied
voltage. We present a detailed discussion of the properties of the
first four cumulants as well as the low and high voltage regimes of
the full counting statistics. The work is an extension of the results
of Pilgram and the author, Phys. Rev. Lett. {\bf 94}, 086806 (2005).
\end{abstract}
\section{Introduction}
\label{intro}
The mechanism for charge transport in voltage biased superconducting
junctions is Multiple Andreev Reflections (MAR)
\cite{KBT82,OTBK83}. Coherent MAR-theory
\cite{CohMAR1,CohMAR2,CohMAR3} has provided a quantum mechanical
picture for current transport in mesoscopic superconducting
junctions. The predictions of the theory were found to be in
impressive agreement with experimental results
\cite{cohexp1,cohexp2,cohexp3}. This development, with the emphasis on
the properties of atomic point contact junctions, was recently
reviewed \cite{MARrew}. Additional insight in coherent MAR transport,
as e.g. the effective charge transferred, has been obtained by
investigations of the current noise \cite{Blantrew}, both theoretical
\cite{CohTheoryNoise1,CohTheoryNoise2,CohTheoryNoise3,CohTheoryNoise4}
and experimental
\cite{CohExpNoise1,CohExpNoise2,CohExpNoise3}. Recently the
theoretical interest has turned to the full distribution of current
fluctuations \cite{Naz}, the Full Counting Statistics (FCS) of
coherent MAR-transport \cite{Johansson03,Cuevas03,Cuevas04}.

In several types of superconducting junctions, e.g. long diffusive
superconducting-normal-superconducting (SNS) junctions, quasiparticles
lose phase coherence when traversing the normal part of the junction
and both the ac and dc Josephson currents are suppressed. This is the
regime of incoherent MAR-transport. The properties of the current in
the incoherent MAR-regime have been thoroughly studied, see
e.g. \cite{Bez} and references therein. In addition, in several works
the current noise has been investigated, both theoretically
\cite{LongTheoryNoise1,LongTheoryNoise2,LongTheoryNoise3} and
experimentally
\cite{LongDiff1,LongDiff2,LongDiff3,LongDiff4}. Recently experimental
data on the third current cumulant in incoherent diffusive SNS
junctions was presented by Reulet et al \cite{Reulet}. Moreover,
Pilgram and the author \cite{PilPRL} presented a general theory for
FCS of incoherent MAR. The theoretical result for a diffusive SNS
junction showed qualitative similarities with the experimental data
\cite{Reulet}. The results in Ref. \cite{PilPRL} were mainly derived
with a stochastic path integral technique \cite{SPI1,SPI2}, however
the extension of the FCS for diffusive SNS to arbitrary voltage bias
was based on a more intuitive picture of incoherent
MAR-transport. Appealing to this intuitive picture, in this work we
first present a detailed derivation and then a thorough analysis of
the FCS of incoherent MAR in diffusive SNS. Moreover, a formal
derivation based on the stochastic path integral approach is presented
in the appendix.

\section{Model and theory}

We consider a junction with a diffusive normal conductor connected to
two superconducting terminals. The superconductors have an energy gap
$\Delta$ and a voltage $V$ is applied between the superconductors. The
phase breaking length in the normal conductor is taken to be much
shorter than the length of the conductor and the transport is
consequently incoherent \cite{incohcond,incohcond2}. We make the
simplifying assumptions that the junction is perfectly voltage biased
and that the scattering is completely elastic. The
normal-superconducting(NS) interfaces are assumed to be
transparent. Due to the diffusive nature of the transport, this gives
that the effective Andreev reflection is perfect inside the
superconducting gap (at quasiparticle energies $|E|<\Delta$) and
negligible outside the gap \cite{NagBut,mycom1}. Throughout the paper
we take the temperature to be well below the gap, $kT \ll \Delta$, and
consider only zero frequency transport properties.

\begin{figure}[h]
\centerline{\psfig{figure=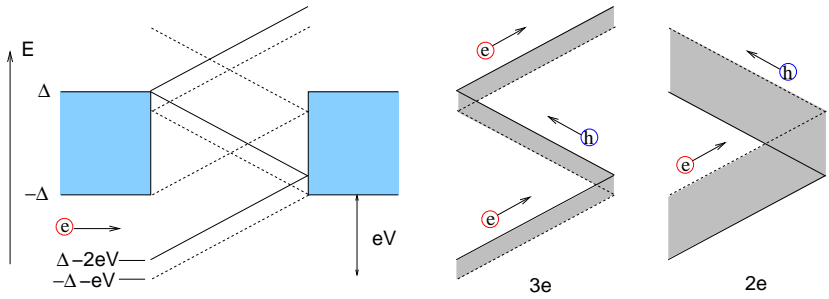,width=0.95\linewidth}}
\caption{Left: MAR-transport in energy space: An electron-like
quasiparticle is injected from the left superconducting terminal at an
energy below the gap, $E<-\Delta$. The quasiparticle propagates via
MAR to an energy above the gap, $E>\Delta$, at which it escapes out
into one of the superconducting terminals. Right: Energy windows for
the transfer of an effective charge $2e$ and $3e$.}
\label{fig1}
\end{figure}

In the paper we will appeal to a qualitative scattering picture of MAR
\cite{CohMAR1,CohMAR2,Shum}, a more formal derivation based on the
approach of Ref. \cite{PilPRL,SPI1,SPI2} is presented in the
appendix. Within the scattering approach, the transport of
quasiparticles in the junction can be described as MAR-transport in
energy space \cite{Johansson}: a quasiparticle below the gap injected
from the superconductor into the normal conductor gain energy by
successive Andreev reflections before escaping out into the
superconductors at energies above the gap. This is illustrated to the
left in fig. \ref{fig1} for an applied voltage $\Delta<eV<2\Delta$.
Due to the perfect effective Andreev reflection at the
normal-superconducting interfaces, there are two possible processes by
which a quasiparticle, injected from the left, can propagate from
filled states below the gap to empty states above the gap.

$\bullet$ for injection energies $-\Delta-eV<E<\Delta-2eV$ a
quasiparticle traverses the gap three times (gaining an energy $3eV$)
as electron $\rightarrow$ hole $\rightarrow$ electron, effectively
transferring a charge $3e$ from left to right.

$\bullet$ for injection energies $\Delta-2eV<E<-\Delta$ a
quasiparticle traverses the gap twice (gaining an energy $2eV$) as
electron $\rightarrow$ hole, effectively transferring a charge $2e$
from left to right.

The two transport processes are shown to the right in
Fig. \ref{fig1}. Importantly, only these processes contribute to the
net charge transport. We emphasize that in the scattering processes
the net charge transferred across the junction is given by the
electron charge multiplied by the net number of quasiparticle
traversals or equivalently, the gained energy divided by the voltage
$V$. This holds independently on the injection energy and the type of
quasiparticle injected \cite{Johansson03}. We also note that for
quasiparticles injected from the right there are, due to the symmetry
of the junction, two equivalent transport processes. Below, we for
simplicity consider only the left injected quasiparticles and multiply
all the derived transport properties with a factor of two.

\subsection{Generalization to arbitrary voltage}

The scenario presented above can be generalized to arbitrary applied
voltage. We have that for a given voltage
\begin{equation}
\frac{2\Delta}{n+1}<eV<\frac{2\Delta}{n}, \hspace{1cm} n=0,1,2,...
\label{volt}
\end{equation}
there are two relevant transport processes for left injected
quasiparticles.

$\bullet$ quasiparticles injected at energies
\begin{equation}
-\Delta-eV<E<\Delta-(n+1)eV
\label{proc1}
\end{equation}
traverse the junction $n+2$ times, transferring a charge $(n+2)e$ from
left to right (gaining an energy $(n+2)eV$).

$\bullet$ quasiparticles injected at energies
\begin{equation}
\Delta-(n+1)eV<E<-\Delta
\label{proc2}
\end{equation}
traverse the junction $n+1$ times, transferring a charge $(n+1)e$ from
left to right (gaining an energy $(n+1)eV$). As emphasized above, no
other processes transferring charge between occupied and unoccupied
states are possible. Consequently, only these two processes
(\ref{proc1}) and (\ref{proc2}) contribute to the net charge
transport.

\subsection{Effective diffusive transport}

\begin{figure}[h]
\centerline{\psfig{figure=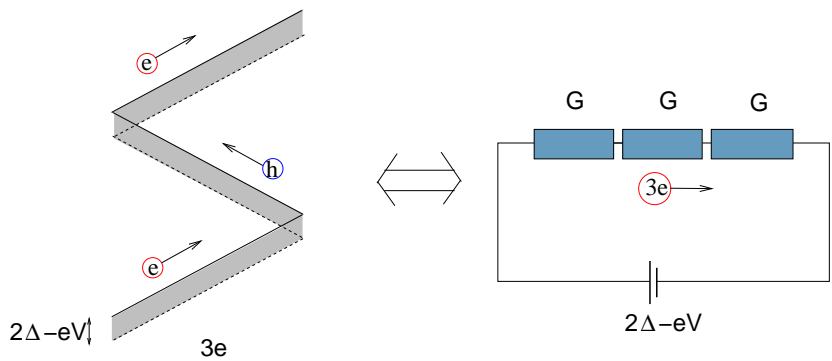,width=0.95\linewidth}}
\caption{Left: MAR-process transporting a charge $3e$ from left to
right. Right: Equivalent normal circuit, conductance $G_3=G/3$, with
quasiparticles with effective charge $3e$.}
\label{fig2}
\end{figure}

The MAR-transport in real space is diffusive. Since the transport is
incoherent, neither the energy of the quasiparticle inside the normal
conductor nor the electron or hole character of the quasiparticle is
of importance for the transport properties. As a consequence, the
process where a quasiparticle traverses the junction a net $n$ number
of times, transferring a net charge $ne$, gives the same (zero
frequency) transport statistics as a process where a quasiparticle
with an effective charge $q_n=ne$ is transported through $n$
normal diffusive conductors in series. The $n$ conductors in series
have a conductance $G_n=G/n$, with $G$ the conductance of the
individual conductors \cite{doublecount}. We point out that this
effective model, illustrated in fig. \ref{fig2}, can be seen as a
straightforward extension and combination of the noise models of
refs. \cite{LongTheoryNoise2,LongTheoryNoise3}.

The ``effective voltage'' across the equivalent normal conductor is
just the energy window available for the process (divided by
$e$). Moreover, each of the two transport processes possible for a
given voltage are independent. Consequently, the the total transport
statistics is obtained by just summing up the contributions from the
two processes. Importantly, since the FCS, or the transport statistics
of diffusive conductors is known \cite{Lee}, the above observation
allows us to directly derive the FCS of the diffusive superconducting
junction.

\section{Cumulants of the current}

Considering only positive voltages, we start with the average
current. The current is the sum of the currents of two transport
processes. For a voltage $2\Delta/(n+1)<eV<2\Delta/n$ we have
\begin{eqnarray}
I(V)&=&V^{(1)}_{n}G_{n+1}\frac{q_{n+1}}{e}+V^{(2)}_{n}G_{n+2}\frac{q_{n+2}}{e} \nonumber \\
&=&\left[(n+1)V-\frac{2\Delta}{e}\right]\frac{G}{n+1}(n+1)+\left[\frac{2\Delta}{e}-nV\right]\frac{G}{n+2}(n+2)\nonumber \\
&=&GV
\label{curr}
\end{eqnarray}
which is just Ohms law. Here, for clarity, we introduced the effective
voltages for the two processes $V^{(1)}_n=V(n+1)-2\Delta/e$ and
$V^{(2)}_n=2\Delta/e-nV$. As is clear from Eq. (\ref{curr}), although
the individual processes depend in a nontrivial way on the applied
voltage via $n$, the total current is just linear in voltage.

\begin{figure}[h]
\centerline{\psfig{figure=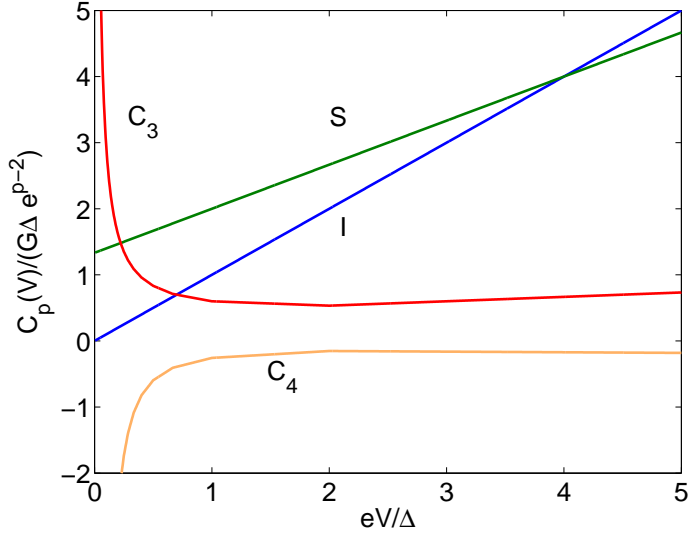,width=0.8\linewidth}}
\caption{The first four cumulants of the current $C_p(V)$ as a
function of voltage. Note that $C_1=I$ and $C_2=S$.}
\label{fig3}
\end{figure}

For the noise, the transport processes are associated with a certain
noise power $P$. For a diffusive conductor the noise power for $n$
conductors in series is $P_n=2eG_n/3$, where the $1/3$ suppression factor
comes from the diffusive nature of the transport
\cite{BeenBut,Nag2}. In addition, the noise is proportional to the
effective charge squared. We can thus write the total noise
\begin{eqnarray}
S(V)&=&V^{(1)}_{n}P_{n+1}\left(\frac{q_{n+1}}{e}\right)^2+V^{(2)}_{n}P_{n+2}\left(\frac{q_{n+2}}{e}\right)^2 \nonumber \\
&=&\frac{2e}{3}\left[(n+1)V-\frac{2\Delta}{e}\right]G_{n+1}(n+1)^2+\left[\frac{2\Delta}{e}-nV\right]G_{n+2}(n+2)^2
\nonumber \\
&=&\frac{2e}{3}G\left[V([n+1]^2-n[n+2])+\frac{2\Delta}{e}([n+2]-[n+1])\right] \nonumber \\
&=&\frac{4\Delta}{3}G\left(1+\frac{eV}{2\Delta}\right)
\end{eqnarray}
which is the known result of
Refs. \cite{LongTheoryNoise2,LongTheoryNoise3}. It is interesting to
note that again the total noise shows no $n$-dependence, it is a
linear function of voltage, with a zero voltage offset. However, going
to the third cumulant $C_3$, the result does depend on $n$, i.e. it
displays a subharmonic gap structure in the form of kinks at voltages
$2\Delta/em$ with $m=1,2,..$. The expression for the third cumulant is
given, along the same lines as for current and noise, by
\begin{eqnarray}
C_3(V)&=&\frac{e^2}{15}G\left[eV(1-n-n^2)+\Delta(6+4n)\right]
\label{C3}
\end{eqnarray}
In the same way we also get the fourth cumulant
\begin{eqnarray}
C_4(V)&=&\frac{e^3}{105}G\left[eV(-1+4n+6n^2+2n^3)-2\Delta(7+9n+3n^2)\right]
\label{C4}
\end{eqnarray}
The first four cumulants are plotted in
fig. \ref{fig3}.
We see that, just as for a normal diffusive conductor, the first three
cumulants are manifestly positive but the fourth cumulant is
negative. In addition, while the first two cumulants are finite at low
voltages, the third and fourth cumulants diverge for $V\rightarrow
0$. Interestingly, in the experimental data \cite{Reulet} the third
cumulant shows a clear increase with decreasing voltage, down to a
cut-off voltage. In this context we note that e.g. inelastic
scattering, not accounted for in our model, will remove the low
voltage divergences, similar to the situation for the noise
\cite{CohTheoryNoise2,LongTheoryNoise2,LongTheoryNoise3}.

\begin{figure}[h]
\centerline{\psfig{figure=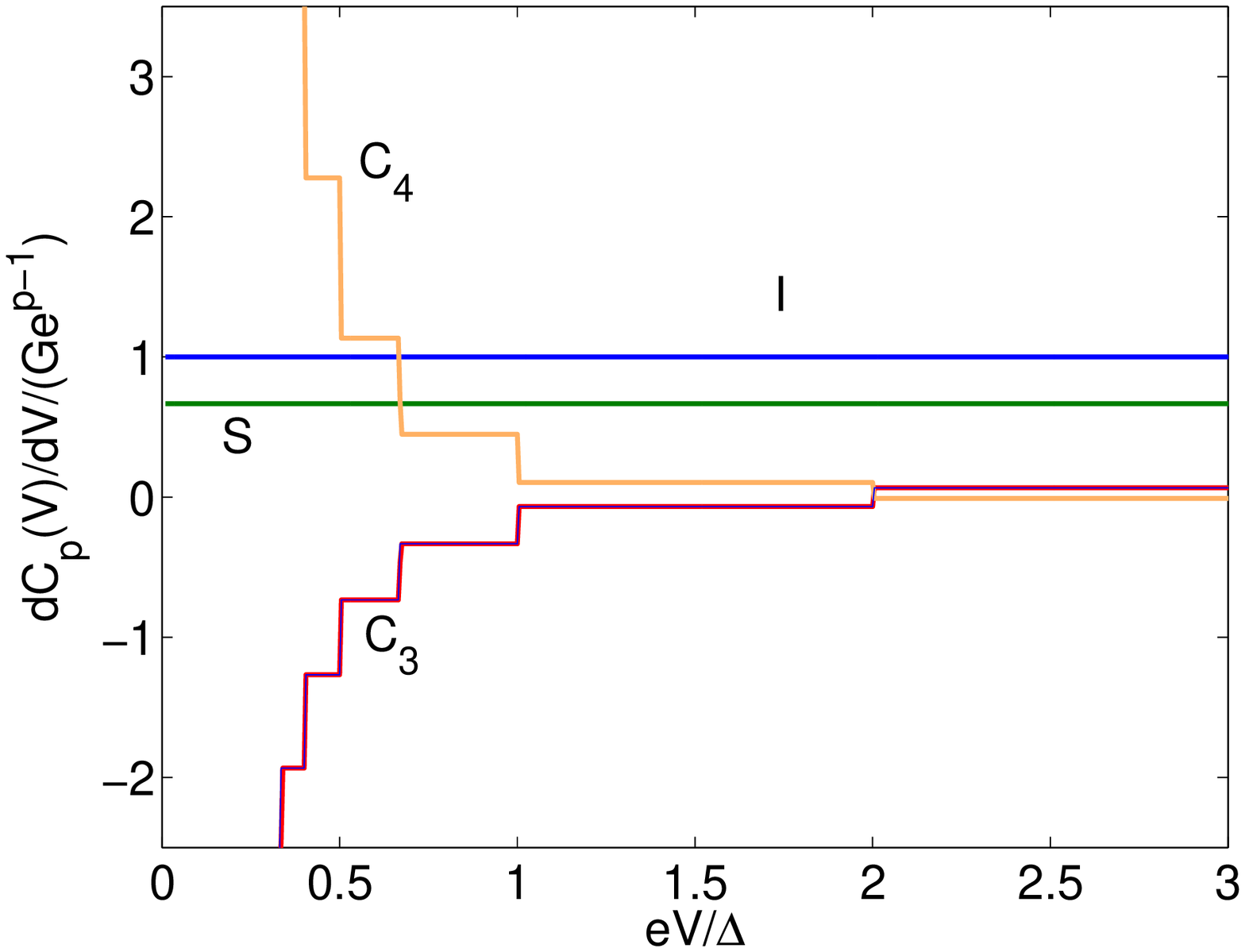,width=0.8\linewidth}}
\caption{The first four differential cumulants $dC_p(V)/dV$ as a function of
voltage. The third and fourth cumulants show a clear step structure.}
\label{fig4}
\end{figure}

The subharmonic gap structure in the voltage dependence of the third
and fourth cumulant is hardly visible in fig. \ref{fig3}. However
turning to the differential cumulants, the derivatives of the
cumulants with respect to the voltage, the subharmonic gap structure
becomes apparent. The differential cumulants are shown in
fig. \ref{fig4}.
We note that the third and the fourth cumulants display a clear step
structure. From Eqs. (\ref{C3}) and (\ref{C4}) we have
\begin{eqnarray}
\frac{dC_3(V)}{dV}&=&\frac{e^3}{15}G(1-n-n^2)
\label{C3x}
\end{eqnarray}
and
\begin{eqnarray}
\frac{dC_4(V)}{dV}&=&\frac{e^4}{105}G(-1+4n+6n^2+2n^3)
\label{C4x}
\end{eqnarray}
We point out that in contrast to the normalized cumulants in coherent
single mode junctions in the tunneling regime
\cite{CohTheoryNoise3,CohTheoryNoise4,Johansson03,Cuevas03,Cuevas04},
the height of the steps of the differential cumulants can not directly
be related to an effective charge. 

\section{Full counting statistics}

With the knowledge of the FCS of a normal diffusive conductor
\cite{Lee}, the scheme above can readily be extended to all higher
cumulants. The cumulant generating function, from which the cumulants
follow by successive derivatives with respect to the counting field
$\chi$, is then given by the sum of the generating functions for the
two MAR-processes
\begin{eqnarray}
F(\chi)&=&\frac{\tau}{e^2}G\left(\frac{(n+1)eV-2\Delta}{n+1}\mbox{asinh}^2\sqrt{e^{\chi (n+1)}-1} \right. \nonumber \\
&&\left. +\frac{2\Delta-neV}{n+2}\mbox{asinh}^2\sqrt{e^{\chi (n+2)}-1}\right).
\label{diffusive}
\end{eqnarray}
where $\tau$ is the measurement time. The current is
$I=(e/\tau)dF/d\chi|_{\chi=0}$, the noise $S=2(e^2/\tau) d^2F/d\chi^2|_{\chi=0}$,
$C_3=(e^3/\tau)d^3F/d\chi^3|_{\chi=0}$ etc. We note that the distribution of
transferred charge $Q$ is given by a Fourier transform
\begin{eqnarray}
P(Q)=\int d\chi e^{F(i\chi)-iQ\chi}
\end{eqnarray}
Here we however do not further investigate the probability
distribution, we instead turn to the low and high voltage limits of
the cumulant generating function.

\section{Voltage limits}

\subsection{Low voltage, $eV \ll \Delta$}
In the limit of low voltage, $eV \ll \Delta$, we have $n \approx 2\Delta/eV$
and the cumulant generating function in Eq. (\ref{diffusive}) is given by
\begin{equation}
F(\chi)=\tau \frac{V^2G}{2\Delta}\mbox{asinh}^2\sqrt{e^{\chi
2\Delta/eV}-1}
\end{equation}
This is just the FCS for a normal diffusive conductor with an
effective charge $e(2\Delta/eV)$ transferred. Importantly, as was
found in Ref. \cite{PilPRL}, this result (with different $G$) holds
for an arbitrary incoherent superconducting junction. The origin of
this universal behavior is that charge transport in incoherent
superconducting junctions at low voltage bias can be described as
diffusion in energy space. This energy diffusion was already noted for
diffusive junctions, in the context of noise, in
Refs. \cite{LongTheoryNoise2,LongTheoryNoise3}.

We note that the different cumulants $C_p$ behave as
\begin{equation}
C_p\propto V^{2-p} 
\end{equation}
This is different from the coherent (short) diffusive SNS-junctions
investigated in Refs. \cite{Averin,Johansson03}, where $C_p\propto
V^{3/2-p}$.

\subsection{High voltage, $eV>2\Delta$}

In the limit of high voltage, $eV>2\Delta$, giving $n=0$ in Eq. (\ref{volt}), the
cumulant generating function in Eq. (\ref{diffusive}) is given by
\begin{equation}
F(\chi)=G\left([eV-2\Delta]~\mbox{asinh}^2\sqrt{e^{\chi}-1}+\Delta~\mbox{asinh}^2\sqrt{e^{2\chi}-1}\right).
\end{equation}
The physical interpretation of this result follows directly from the
model: the first term is due to single particle transport ($\chi$ in
the exponent) with an onset at $eV=2\Delta$ while the second term is
two-particle transport ($2\chi$ in the exponent). The ``excess''
fluctuations, the difference between the cumulant generating functions
for the diffusive junction with the terminals in the superconducting
and normal state respectively, is thus given by
\begin{equation}
F^{exc}(\chi)=F(\chi)-F_N(\chi)=G\Delta\left(\mbox{asinh}^2\sqrt{e^{2\chi}-1}-2\mbox{asinh}^2\sqrt{e^{\chi}-1}\right).
\end{equation}
We note that this result is independent on voltage. It gives the
excess cumulants $I^{exc}=0$, $S^{exc}=4\Delta G/3$,
$C_3^{exc}=2e\Delta G/5$, etc.

\section{Conclusions}

In conclusion, we have investigated the full counting statistics of
incoherent multiple Andreev reflections in diffusive SNS-junctions. An
expression for the cumulant generating function has been derived. A
careful analysis of the first four cumulants and the low and high
voltage limits of the cumulant generating function has been
presented. The results are valid under the assumptions of perfect
voltage bias and the absence of inelastic scattering. To allow for a
quantitative comparison with experiments for the full range of applied
voltages, a theoretical model where these assumptions can be relaxed
would be of great interest.

\section{Acknowledgments} The author thanks Vitaly Shumeiko and Wolfgang Belzig for important comments on the manuscript. The work was supported by the Swedish VR.

\section{Appendix}

\begin{figure}[h]
\centerline{\psfig{figure=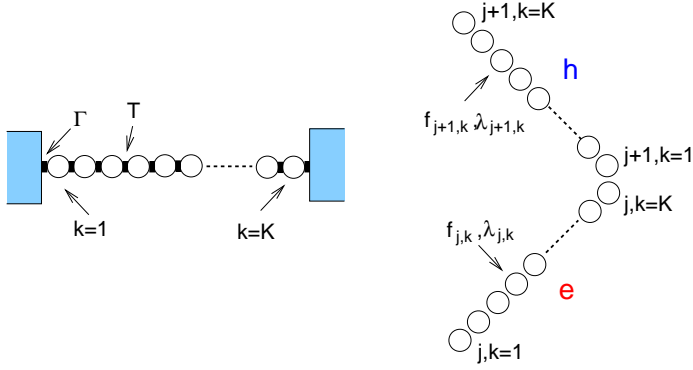,width=0.8\linewidth}}
\caption{Left: discretization of the diffusive wire into $K$
nodes. The contact between the nodes has transparency $T$ and the contact
between the nodes and the superconducting electrodes has transparency
$\Gamma$. Right: the node structure along a part of the
MAR-ladder. The distribution functions $f_{j,k}$ and counting fields
$\lambda_{j,k}$ of the nodes labeled with $j,k$ are shown.}
\label{figapp}
\end{figure}

The cumulant generating function in Eq. (\ref{diffusive}) can be
conveniently derived in a more formal way using the stochastic path
integral approach \cite{SPI1,SPI2} applied to incoherent MAR
\cite{PilPRL}. We consider for simplicity a wire geometry, length $L$,
of the diffusive conductor (for a more general diffusive geometry, see
Ref. \cite{SPI2}). The starting point is to discretize the diffusive
wire into $K\gg 1$ nodes, contacted via quantum point contacts
supporting $M\gg 1$ transport modes with transparency $T$. This
discretization is shown in Fig. \ref{figapp}.
The transparency $T$ and the number of transport modes $M$ are chosen
to give a conductance of the $K$ nodes in series $[MT/K]2e^2/h$ equal
to the conductance G of the wire.

The resulting node structure in energy space, along the MAR-ladder, is
also shown in Fig. \ref{figapp}. Each node is characterized by a
quasiparticle distribution function $f_{j,k}$ and a quasiparticle
counting field $\lambda_{j,k}$, where $1\leq k \leq K$ is the number
of the node and $j$ denotes the rung of the MAR-ladder. The $j$:th
rung goes between NS-interfaces at energies $E_{j-1}$ and $E_{j}$
where $E_j=E+jeV$. We take $j=0$ for energies just below the gap
$-\Delta<E<-\Delta-2eV$, on the left side. The quasiparticles are thus
electrons for rungs with $j$ odd and holes for $j$ even. Note also
that $k$ is counted along the ladder, upwards in energy.

Following the scheme for FCS of incoherent MAR in Ref. \cite{PilPRL},
we write the total generating function as a sum (energy integral) over
the generating functions for different MAR-ladders, which
contribute independently to the total generating function. This gives
\begin{equation}
S[\chi,\{f\},\{\lambda\}]=\int_{-\Delta-2eV}^{-\Delta}dE\sum_{j=-\infty}^{\infty}\left[S^{NS}_{j}+\sum_{k=1}^{K-1}S^N_{j,k}\right]
\label{action}
\end{equation}
where the generating function for a given MAR-ladder is a sum of the
generating functions $S_{j,k}^N$ and $S_{j}^{NS}$ along the
ladder. Here
\begin{eqnarray}
S^N_{j,k}&=&\frac{M\tau}{h}\ln\left[f_{j,k}f_{j,k+1}+f_{j,k}(1-f_{j,k+1})\left[Te^{\lambda_{j,k+1}-\lambda_{j,k}}+R\right]
\right. \nonumber \\ &+&
\left. f_{j,k+1}(1-f_{j,k})\left[Te^{\lambda_{j,k}-\lambda_{j,k+1}}+R\right]
\right. \nonumber \\ &+& \left. (1-f_{j,k})(1-f_{j,k+1})\right]
\label{actionN}
\end{eqnarray}
is the generating function \cite{Levitov} for the quantum point
contact between nodes in the wire, with $R=1-T$. The generating
function $S_{j}^{NS}$ describes the connection between nodes at the
NS-interfaces \cite{Muzy,Belzig}. At energies inside the gap,
$|E_j|<\Delta$, we have
\begin{eqnarray}
S^{NS}_{j}&=&\frac{M\tau}{h}\ln\left[f_{j,K}f_{j+1,1}+f_{j,K}(1-f_{j+1,1})\left[R_Ae^{2\chi_j+\lambda_{j+1,1}-\lambda_{j,K}}+R_N\right]
\right. \nonumber \\ &+&
\left. f_{j+1,1}(1-f_{j,K})\left[R_Ae^{-2\chi_j+\lambda_{j,K}-\lambda_{j+1,1}}+R_N\right]\right. \nonumber
\\ &+& \left. (1-f_{j,K})(1-f_{j+1,1})\right]
\label{NSaction1}
\end{eqnarray}
with $R_A=R_A(E,\Gamma)$ the Andreev reflection probability and
$R_N=R_N(E,\Gamma)=1-R_A$ the normal reflection probability, where
$\Gamma$ is the (normal) transparency of NS-interface. The counting
field $\chi_j$ is taken equal to $\chi$ for $j$ odd and zero for $j$
even, i.e. the transferred charge is counted on the right side of the
junction. It is important to note that $\chi$ counts electrical charge
while $\lambda_{j,k}$ counts quasiparticles.

Outside the gap, at $|E_j|>\Delta$, the Andreev reflection can
effectively be neglected (as discussed above) and we can for
simplicity consider a generating function for a perfect normal
interface, i.e. a NS-interface with the superconductor taken in the
normal state. This gives
\begin{eqnarray}
S^{NS}_{j}&=&\frac{M\tau}{h}\ln\left[f_{j,K}f_j+f_{j,K}(1-f_{j})\left[e^{\chi_j-\lambda_{j,K}}+1\right]
\right. \nonumber \\ &+&
\left. f_{j}(1-f_{j,K})\left[e^{-\chi_j+\lambda_{j,K}}+1\right]\right. \nonumber
\\ &+& \left. (1-f_{j,K})(1-f_{j})\right]
\label{NSaction2}
\end{eqnarray}
above the gap, $E_j>\Delta$ and similarity below the gap. Here
$f_j=f(E_j)$, with $f(E)$ the quasipartice distribution function of
the superconducting electrodes. For $kT \ll \Delta$ considered here,
we have $f(E)=1$ below the gap and $f(E)=0$ above the gap.

To evaluate the generating functions for the ladders we first perform
a transformation
\begin{eqnarray}
\lambda_{j,k} &\rightarrow& \lambda_{j,k}-j\chi, \hspace{0.5cm} j=2,4,6... \nonumber \\
\lambda_{j,k} &\rightarrow& \lambda_{j,k}-(j-1)\chi, \hspace{0.5cm} j=1,3,5....
\label{transf}
\end{eqnarray}
[Under the stated conditions, only $\lambda_{j,k}$ with $j\geq 1$ are
of interest]. This transformation removes the counting field $\chi$
from the generating functions of the NS-interfaces at energies inside
the gap, in Eq. (\ref{NSaction1}). In addition, the generating
function for the NS-interface above the gap, Eq. (\ref{NSaction2}) is
modified as $\chi_j \rightarrow j\chi$. This means that the counting
field $\chi$ is transferred away from the interior of the ladder to
the upper boundary.

Second, we the note that the structure of the generating functions for
the quantum point contact between the nodes in the wire,
Eq. (\ref{actionN}), and for the NS-interface at energies inside the
gap, Eq. (\ref{NSaction1}), are formally identical when $\chi$ is
transferred away (however with $R_A$ instead of $T$). The NS-interface
thus effectively connects the end-nodes $j,K$ and $j+1,1$ of the two
rungs $j$ and $j+1$.

In addition, we emphasize that the boundary conditions (in the
superconductors) for the distribution functions as well as the
counting fields [after the transformation in Eq. (\ref{transf})] are
independent on the electron or hole character of the quasiparticles,
they only depend on the energy $E_j$. Consequently, the fact that
distribution functions and counting fields describe electrons for $j$
odd and holes for $j$ even becomes irrelevant for the charge transfer
and one thus only need to consider the position $j,k$ of the node in
the MAR-ladder.

Taken together, the MAR ladder can thus be represented as a series of
$nK$ nodes, with $n$ equal to the number of rungs on the ladder, or
equivalently the net number of absorbed energy quanta $eV$ when
traversing the junction. Note that since the resistance of the
NS-interface is much smaller than the resistance of the diffusive
wire, the details of the connection at the NS-interface ($R_A$ instead
of $T$) can be neglected. The only difference from a discretized
normal diffusive wire of length $nL$ is then the boundary condition
for the counting field at the end of the wire, being $n\chi$ instead
of just $\chi$ for a normal wire.

Having performed this mapping to a discretized normal wire with
renormalized boundary conditions we can continue with the standard
procedure \cite{SPI2} to derive the total generating function in
Eq. (\ref{action}). We first denote the nodes with a single index $k$
running from $1$ to $nK$ (dropping the rung index). In the limit $K
\gg 1$ the difference between distribution functions as well as
counting fields of two neighboring nodes $k$ and $k+1$ is of the order
$1/K$ and we can expand $f_{k+1}=f_k+\Delta f_k$ and
$\lambda_{k+1}=\lambda_k+\Delta \lambda_k$. Inserting this expansion
into the generating function in Eq. (\ref{actionN}) we have to leading
order in $\Delta f_k$ and $\Delta \lambda_k$
\begin{equation}
S_k^N=\frac{M\tau}{h}T\left[\Delta\lambda_k\Delta f_k-f_k(1-f_k)\Delta \lambda_k^2\right]
\end{equation}
Taking the continuum limit and summing over all generating functions
for different $k$ we get the generating function for a given energy
$E$ as \cite{SPI2}
\begin{equation}
S[\chi,f,\lambda]=\frac{M\tau}{h}\frac{T}{Kn}\int_0^{1}dx \left[\frac{d\lambda}{dx}\frac{df}{dx}-f(1-f)\left(\frac{d\lambda}{dx}\right)^2\right] 
\label{genfcnwire}
\end{equation}
where we have changed from summation over nodes to an integral over
the dimensionless position $0<x<1$ along the wire, i.e $f=f(x)$ and
$\lambda=\lambda(x)$.

Second, this generating function is varied over all possible
configurations of $f$ and $\lambda$. This results in a stochastic path
integral formulation for the averaged generating function $F(\chi)$
\begin{equation}
\exp(F[i\chi])=\int {\mathcal D}f{\mathcal D}\lambda \exp(S[i\chi,f,i\lambda])
\end{equation}
Under the semiclassical conditions considered here, this path integral
can be solved in the saddle point approximation. The saddle point
equations are given by the functional derivatives
\begin{eqnarray}
\frac{\delta S[\chi,f,\lambda]}{\delta \lambda}&=&2\frac{d}{dx}\left[\frac{d\lambda}{dx}f(1-f)\right]-\frac{d^2f}{dx^2}=0 \nonumber \\
\frac{\delta S[\chi,f,\lambda]}{\delta f}&=&(1-2f)\left(\frac{d\lambda}{dx}\right)^2+\frac{d^2\lambda}{dx^2}=0
\end{eqnarray}
Solving these equations for $\lambda$ and $f$ in terms of $\chi$,
inserting the solutions back into the generating function and
performing the integration over $x$ then yields (see Ref. \cite{SPI2}
for details)
\begin{equation}
F(\chi)=\frac{\tau}{e^2}\frac{G}{n}\mbox{asinh}^2\sqrt{e^{n\chi}-1}
\end{equation}
describing the FCS of a diffusive wire with conductance $G_n=G/n$ and
a renormalized charge $e \rightarrow ne$. Finally, performing the
integral over energy in Eq. (\ref{action}), taking into account energy
window for the two MAR-ladders, gives the generating function in
Eq. (\ref{diffusive}). This concludes the formal derivation.


\begin{thebibliography}{}
\bibitem{incohcond2} In the incoherent regime considered the proximity effect is completely suppressed. The effect of a finite proximity effect close
to the NS-interfaces was investigated in Ref. \cite{Bez} in the
context of MAR.
\bibitem{doublecount} Note that in the scattering approach to
superconducting systems \cite{Datta,Shum} electrons and holes, instead
of spin up and spin down, are counted. As a consequence, the analogy
between MAR-transport and normal transport gives the single spin
transport statistics. A factor of two comes from counting also the
right injected quasiparticles, as discussed in the text.
\end{thebibliography}
\end{document}